\documentclass[useAMS,usenatbib]{mn2e}
\usepackage{graphicx}
\usepackage{amsmath}

\def\be{\begin{equation}}
\def\ee{\end{equation}}
\def\ba{\begin{eqnarray}}
\def\ea{\end{eqnarray}}
\def\go{\mathrel{\raise.3ex\hbox{$>$}\mkern-14mu
             \lower0.6ex\hbox{$\sim$}}}
\def\lo{\mathrel{\raise.3ex\hbox{$<$}\mkern-14mu
             \lower0.6ex\hbox{$\sim$}}}

\def\bxi{{\mbox{\boldmath $\xi$}}}
\def\br{{\bf r}}
\def\bOmega{{\bf \Omega}}

\def\cD{{\cal D}}
\def\mmp{{mm'}}
\def\sinT{\sin\!\Theta}
\def\cosT{\cos\!\Theta}
\def\cU{{\cal U}}
\def\tomega{{\tilde\omega}}
\def\ct{{\rm c}_\theta}
\def\st{{\rm s}_\theta}
\def\bm{{\bar m}}

\def\bk{{\bf k}}

\begin{document}
\title[Tidal Dissipation in Planet-Hosting Stars]
{Tidal Dissipation in Planet-Hosting Stars:
Damping of Spin-Orbit Misalignment and Survival of Hot Jupiters}
\author[D.~Lai]
{Dong Lai\thanks{Email: dong@astro.cornell.edu}\\
Center for Space Research, Department of Astronomy, 
Cornell University, Ithaca, NY 14853, USA\\}

\pagerange{\pageref{firstpage}--\pageref{lastpage}} \pubyear{2011}

\label{firstpage}
\maketitle

\begin{abstract}
Observations of hot Jupiters around solar-type stars with very short
orbital periods ($\sim 1$~day) suggest that tidal dissipation in such
stars is not too efficient so that these planets can survive against
rapid orbital decay. This is consistent with recent theoretical works,
which indicate that the tidal quality factor, $Q_\star$, of
planet-hosting stars can indeed be much larger than the values
inferred from the circularization of stellar binaries. On the other
hand, recent measurements of Rossiter-McLaughlin effects in transiting
hot Jupiter systems not only reveal that many such systems have
misaligned stellar spin with respect to the orbital angular momentum
axis, but also show that systems with cooler host stars tend to have
aligned spin and orbital axes. Winn et al. suggested that this
obliquity - temperature correlation may be explained by efficient
damping of stellar obliquity due to tidal dissipation in the
convection zone of the star. This explanation, however, is in apparent
contradiction with the survival of these short-period hot Jupiters. We
show that in the solar-type parent stars of close-in exoplanetary
systems, the effective tidal $Q_\star$ governing the damping of
stellar obliquity can be much smaller than that governing orbital
decay. This is because for misaligned systems, the tidal potential
contains a Fourier component with frequency equal to the stellar spin
frequency (in the rotating frame of the star) and rotating opposite to
the stellar spin. This component can excite inertial waves in the
convective envelope of the star, and the dissipation of inertial waves
then leads to a spin-orbit alignment torque and a spin-down torque,
but not orbital decay. By contrast, for aligned systems, such inertial
wave excitation is forbidden since the tidal forcing frequency is much
larger than the stellar spin frequency.  We derive a general effective
tidal evolution theory for misaligned binaries, taking account of
different tidal responses and dissipation rates for different tidal
forcing components.
\end{abstract}


\section{Introduction}

\subsection{Tidal Dissipation in Planet-Hosting Stars: A Conundrum}

Recent measurements of stellar obliquity in transiting
exoplanetary systems using the Rossiter-McLaughlin effect
have shown that a significant fraction of 
hot Jupiter systems have misaligned stellar spin with respect to the 
planetary angular momentum axis (e.g., H\'ebrard et al.~2008; 
Winn et al.~2009,2010,2011; Johnson et al.~2009; Narita et al.~2009;
Triaud et al. 2010; Pont et al.~2010; Simpson et al.~2011; Moutou et al.~2011).
%
This suggests that a large population hot Jupiters
are formed through dynamical planet-planet scatterings
(e.g., Rasio \& Ford 1996; Weidenschilling \& Marzari 1996;
Zhou et al.~2007; Chatterjee et al.~2008; Juric \& Tremaine 2008)
and more importantly, through 
secular interactions between multiple planets or
Kozai effect induced by a distant companion (e.g., Wu \& Murray 2003;
Fabrycky \& Tremaine 2007;
Nagasawa et al.~2008; Naoz et al. 2011a,b;
Wu \& Lithwick 2011; Katz et al.~2011), 
although other effects involving star-disc
interactions (Lai et al.~2011; Foucart \& Lai 2011) and the 
assembly of protoplanetary discs (Bate et al.~2010; Thies et al.~2011)
may also play a role in producing spin-orbit misalignment.

Recent observations have also revealed an intriguing correlation
between stellar obliquity $\Theta$ and effective temperature $T_{\rm eff}$:
Winn et al.~(2010) found that the misaligned systems tend to have
hotter parent stars ($T_{\rm eff}\go 6250$~K, corresponding to stellar
mass $M_\star\go 1.3M_\odot$), while the systems containing cooler
stars have small obliquities. This trend was also found from 
a recent analysis of the stellar rotation velocities (Schlaufman 2010).
Although this $\Theta$--$T_{\rm eff}$ 
correlation may indicate different planet formation
mechanisms for stars of different masses, Winn et al.~(2010) pointed
out that $T_{\rm eff} = 6250$~K corresponds to the temperature below
which stars contain a large sub-surface convective envelope, and suggested
that tidal dissipation in 
these stars lead to the damping of their obliquities. Indeed, the
facts that effective temperature is more closely related to obliquity
than stellar mass and that a few low-mass and long-period planets are
exceptions to the $\Theta-T_{\rm eff}$ correlation (Winn et
al.~2010,2011), support the idea of tidal damping of spin-orbit
misalignment in solar-type stars.
Most recently, Triaud (2011) found a correlation between the obliquity
and stellar age, suggesting a tidal alignment timescale of about 2.5~Gyr.


It has long been recognized that tidal dissipation in planet-hosting
stars could play an important role the evolution and
survival of hot Jupiter systems (e.g., Lin et al.~1996;
Rasio et al.~1996; Marcy et al.~1997; 
Sasselov 2003; Dobbs-Dixon et al.~2004; 
Barker \& Ogilvie 2009; Jackson et al.~2009;
Levrard et al.~2009; Leconte et al.~2010; Matsumura et al.~2010).
The strength of tidal
dissipation is usually parameterized by a dimensionless quality factor
$Q_\star$.  When the orbital mean motion $\Omega$ 
is larger than the stellar spin
frequency $\Omega_s$, the orbital decay timescale (for circular orbits)
is 
\ba &&t_a=\left|{a\over \dot a}\right|={2Q_\star'\over 9}
\left(\!{M_\star\over M_p}\!\right) \left(\!{a\over
  R_\star}\!\right)^{\!5}{1\over \Omega}\nonumber\\ 
&&\quad \simeq
1.28\,\left({Q_\star'\over 10^7}\right) \!\left(\!{M_\star\over
  10^3M_p}\!\right)
\!\left(\!{\bar\rho_\star\over\bar\rho_\odot}\!\right)^{\!\!5/3}
\!\left(\!{P\over 1\,{\rm d}}\right)^{\!\!13/3}{\rm Gyr}, 
\ea 
where $Q_\star'=3Q_\star/(2k_2)$ is the reduced tidal quality factor,
$k_2$ is the Love number, and $\bar\rho_\star$ is the mean density of 
the star and $\bar\rho_\odot$ is the solar value.
The inspiral time for a planet into its host
star is $(2/13)t_a$. The observations of hot Jupiters with the
shortest orbital periods (such as 
WASP-18b, 0.94~d; WASP-19b, 0.79~d; and WASP-43b, 0.81~d;
Hellier et al.~2009,2011; Hebb et al.~2010) 
suggest $Q_\star'\go 10^8$ (see Brown et al.~2011).  
Such a large $Q_\star'$ value is consistent with 
recent theoretical works (see Sec.~1.2) on the physics
of tidal dissipation in planet-hosting solar-type stars
(Ogilvie \& Lin 2007; Barker \& Ogilvie 2010,2011; Penev \& Sasselov 2011).
On the other hand, in the often-used tidal evolution equations
(see Sec.~2.5), the damping time for stellar obliquity $\Theta$ is 
\ba
&&t_\Theta=\left|{\sin\Theta\over\dot\Theta}\right|\simeq
\left({2S\over L}\right)t_a\nonumber\\ &&\quad \simeq
1.13\,t_a\left({\kappa\over 0.1}\right) \!\left(\!{M_\star\over
  10^3M_p}\!\right)\!
\!\left(\!{\bar\rho_\odot\over\bar\rho_\star}\!\right)^{\!\!2/3}
\!\!\left(\!{10\,{\rm d}\over P_s}\!\right) \!\left(\!{1\,{\rm d}\over
  P}\right)^{\!\! 1/3}\!\!, 
\label{eq:tTheta}\ea 
where $S$ and $L$ are the stellar spin and orbital angular momenta,
respectively, $P_s$ is the spin period and the
moment of inertia of star is $\kappa MR^2$. Thus, for typical
parameters of hot Jupiter systems, $t_\Theta\sim t_a$, and a tidal quality
factor $Q_\star'\go 10^8$ would not cause significant damping of
$\Theta$. To put it in another way, a reduction in the stellar obliquity is
accompanied by a similar amount of orbital decay, $\Delta a/a\simeq
(2S/L)\Delta\Theta/\sin\!\Theta$. This poses a severe problem to the
(otherwise appealing) tidal damping interpretation of the stellar 
obliquity -- effective temperature correlation (Winn et al.~2010). 
One way out of this problem is to assume that the star's convective 
envelope is weakly coupled to its radiative core, thus reducing 
the obliquity damping time for the stellar envelope
(Winn et al.~2010). This assumption, however, is difficult to 
substantiate, as fluid instabilities may develop in the presence of
large differential rotation (especially when the directions of 
rotation vary across the star) to quickly couple the rotations of
the core and the envelope.

To recapitulate, there is a conundrum concerning the efficiency of
tidal dissipation in planet-hosting solar-type stars: On the one hand,
the survival of hot Jupiters with shortest orbital periods and recent
theoretical works both indicate that stellar tidal dissipation induces
only modest or negligible orbital decay. On the other hand, the
observed stellar obliquity - effective temperature correlation
suggests that tidal dissipation is important in damping stellar
obliquity

In this paper, we show that tidal damping of spin-orbit misalignment
can be much more efficient than tidal damping of the orbit. In
another word, the effective tidal quality factor for the former
process can be much smaller than the latter. This provides a natural
resolution to the conundrum discussed above.


\subsection{Basic Idea}

Many previous works on tidal evolution in hot Jupiter systems (e.g.,
Rasio et al.~1996; Sasselov 2003; Dobbs-Dixon et al.~2004; Barker \&
Ogilvie 2009; Jackson et al.~2009; Levrard et al.~2009; 
Hansen 2010; Matsumura et al.~2010) were based on the weak friction theory of
equilibrium tides. This theory considers large-scale
quadrupole distortion of the star, and parameterizes tidal
dissipation by a dimensionless quality factor $Q_\star$ or more generally,
by a constant tidal lag time $\Delta t_L$. 
The theory was first formulated by Darwin (1880), and
extensively applied to solar-system bodies (e.g., Goldreich \& Solter 1966) 
and stellar binaries (see Zahn 2008 for a review). These applications
have proved very useful since they provide empirical estimates or
constraints on the values of $Q_\star$ for various systems.
The most general (arbitrary orbital eccentricity and spin-orbit inclination 
angle) and correct equations for tidal evolution based on this 
theory were derived by Alexander (1973), and were also elaborated by others
(e.g., Hut 1981; Eggleton et al.~1998; Correia \& Laskar 2010).

Although it is well recognized that the equilibrium tide theory is
a parameterized theory, with all the physics of tidal dissipation hidden
in a single parameter $Q_\star$ or $\Delta t_L$ (the Love number of the body 
can be absorbed into the definition of these parameters), it is not widely
appreciated that the effective tidal $Q_\star$ for different processes
(e.g., spin-orbit alignment and orbital decay) can be different.  
In another word, the widely-used tidal evolution equations based on 
equilibrium tide theory can be
incorrect even at the parameterized level. Indeed, the conundrum discussed
in Sect.~1.1 arises because Eq.~(\ref{eq:tTheta}) assumes that 
the tidal $Q_\star$ for stellar obliquity damping is similar to the $Q_\star$ 
for orbital decay. In fact, this is incorrect, as we explain below.

There are three channels of tidal dissipations in solar-type stars:

(i) Equilibrium tides. The large-scale quasi-static tidal 
bulge can be damped by turbulent viscosity in 
the star's convective envelope (Zahn 1977,1989). 
The major uncertainty involves
how the effective viscosity derived from the mixing-length theory,
$\sim v_t l_t/3$ (where $v_t$ and $l_t$ are the velocity and size of
convective eddies, respectively), is reduced when the tidal forcing period
$P_{\rm tide}$ is shorter than the convective turnover time 
$\tau_t=l_t/v_t$ (see Goodman \& Oh 1997). 
Recent simulations (Penev et al.~2009,2011)
suggests that the reduction factor is about $P_{\rm tide}/(2\pi\tau_t)$ 
(for a limited range of $P_{\rm tide}$), and the corresponding 
tidal $Q_\star$ well exceeds $10^8$ (Penev \& Sasselov 2011).
An even larger $Q_\star$ will result if the reduction factor 
$(P_{\rm tide}/2\pi\tau_t)^2$ is used (see Ogilvie \& Lin 2007).

(ii) Excitation and damping of internal gravity waves 
(Goodman \& Dickson 1998; Ogilvie \& Lin 2007; Barker \& Ogilvie 2010,2011).
These waves (also called Hough waves when modified by rotation) are
launched at the bottom of the star's convective envelope and propagate
toward the stellar center. If they attain sufficient amplitudes at the center,
wave breaking will occur; this will produce significant tidal
dissipation, corresponding to $Q_\star \sim {\rm a~few}\times 
10^5(P/1\,{\rm day})^{8/3}$ [assuming the orbital period $P$ is much 
shorter than the spin period; see Barker \& Ogilvie (2010)].
If the waves are reflected coherently at the stellar center (e.g., by a 
small convective core) before nonlinear breaking (Terquem et al.~1998), 
only weak dissipation will result ($Q_\star\go 10^8$). 
The latest calculations by Barker \& Ogilvie (2010,2011) suggest that 
while the nonlinear wave breaking certainly occurs for binary
stars, it is probably unimportant for exoplanetary systems 
-- this would explain the survival of short-period hot Jupiters against
orbital decay (see Weinberg et al.~2011).

(iii) Excitation and damping of inertial waves.  Recent theoretical
works on dynamical tides in rotating planets (Ogilvie \& Lin
2004,2007; Ogilvie 2005,2009; Wu 2005a,b; Papaloizou \& Ivanov 2005;
Goodman \& Lackner 2009) and stars (Savonije \& Papaloizou 1997;
Papaloizou \& Savoniji 1997; Savonije \& Witte 2002);
Ogilvie \& Lin 2007) have
emphasized the importance of inertial waves driven by Coriolis force.
When the tidal forcing frequency (in the rotating frame) $\tomega$ is
less than twice the spin frequency ($\Omega_s$), short-wavelength inertial
waves can be excited. In particular, when these waves are confined to
a spherical shell (as in the convection zone outside the solid core of
a giant planet or in the convective envelope of a solar-type star),
tidal disturbances are concentrated in very narrow regions (called
``wave attractors'') where dissipation takes place (Ogilvie \& Lin
2004; Ogilvie 2009; Goodman \& Lackner 2009).  It appears
that this mechanism can explain the tidal $Q$ ($\sim 10^6$) of giant
planets and, when combined with internal gravity wave damping [see (ii)
above], can also explain the dissipation required for the
circularization of stellar binaries. However, for solar-type stars in hot
Jupiter systems, the inertial wave dissipation mechanism is 
not expected to operate, since the tidal frequency 
$\tomega=2(\Omega-\Omega_s)$ (assuming circular orbit
and aligned stellar spin) is larger than $2\Omega_s$ for typical 
parameters (e.g., $P_s\sim 10$~d and $P\sim 1$~d).

The main point of our paper concerns inertial wave dissipation in
the parent stars of hot Jupiter systems when the stellar spin 
${\bf S}$ is misaligned with the orbital angular momentum ${\bf L}$.  For
a circular binary, in the inertial coordinate system with the $Z$-axis
along ${\bf L}$, the tidal potential has two components (to the
quadrupole order), with frequencies
$\omega_{m'}=m'\Omega$ (where $m'=0,2$). As seen in the rotating frame of the 
star, the tidal frequencies become 
\be
\tomega_{mm'}=m'\Omega-m\Omega_s,
\label{eq:tomega}\ee
with $m=0,\pm 1,\pm 2$. For an aligned system, only the $m=m'=2$ component
of the tidal potential is nonzero and involved in tidal dissipation.
(The $m=m'=0$ component is also nonzero, but it does not transfer 
energy and angular momentum since it is completely static.)
For misaligned systems, however, all seven tidal components [all combinations
of $(m,m')$ except $m=m'=0$; note that for $m'=0$, the $m=-1(-2)$ component is 
physically identical to the $m=1(2)$ component] 
contribute to the transfers of tidal energy and/or angular momentum. 
In general, each of the 7 distinct 
components will generate tidal disturbance with its own quality factor
$Q_{mm'}$. Of particular interest is the $(m,m')=(\pm 1,0)$ components. They 
have $\tomega_{mm'}=\mp \Omega_s$ and the angular pattern 
frequency is $\tomega/m=-\Omega_s$ (the negative sign 
means it is retrograde with respect to the spin). They can generate inertial
waves in the convection zone, 
and therefore making a significant or dominant contribution 
to the tidal alignment of spin-orbit inclination. These components,
however, do not contribute to orbital decay since they are static 
in the inertial frame and do not transfer energy (see Sect.~3).

In summary, while for aligned hot Jupiter systems, the effective stellar
tidal $Q_\star$ governing orbital decay (and contributing to 
orbital circularization\footnote{Tidal dissipation in the planet also 
contributes to circularization.} may be quite large ($\go 10^8$) 
because channel (i) and channel (ii)
(i.e., equilibrium tides and gravity waves)
are ineffective and channel (iii) (inertial waves) is forbidden,
for misaligned systems, inertial wave excitation becomes possible,
which may provide an efficient damping mechanism for the spin-orbit misalignment.

The remainder of this paper is organized as follows. In Sect.~2
we develop a general effective theory for tidal evolution of misaligned
binaries. It is ``general'', because the theory takes account of the
different responses (including both equilibrium and dynamical tides)
of the star to different frequency components of the tidal potential.
It is ``effective'', because the responses and dissipations of different
components are treated in a parameterized way.
In Sect.~3 we present the tidal evolution equations due to 
inertial wave dissipation for misaligned hot Jupiter systems.
We conclude in Sect.~4.

\section{Tidal Evolution of Misaligned Binaries: A General 
Effective Theory}
\label{sec:general}

\subsection{Tidal Potential}

We consider a star of mass $M$, radius $R$ and spin 
$\Omega_s$ (along the direction ${\hat{\bf S}}$) orbited by 
a companion (planet) of mass $M'$.
The orbital semi-major axis is $a$ and the orbital angular frequency is 
$\Omega$. We allow for general spin-orbit inclination angle 
$\Theta$ (the angle between the spin angular momentum ${\bf S}=S{\hat{\bf S}}$ 
and the orbital angular momentum ${\bf L}=L{\hat{\bf L}}$), 
but consider circular orbit 
for simplicity. In the spherical coordinate system centered on $M$
with the $Z$-axis along ${\hat{\bf L}}$, the tidal potential produced
by $M'$ can be expanded in terms of spherical harmonics:
\be
U(\br,t)=
-GM'\sum_{m'}{W_{2m'}r^2\over a^3}\,e^{-im'\Omega t}
Y_{2m'}(\theta_L,\phi_L), \label{eq:potential}
\ee
where $m'=0,\pm 2$, with 
$W_{20}=-(\pi/5)^{1/2}$ and $W_{2\pm 2}=(3\pi/10)^{1/2}$.
To study the dynamical response of stellar fluid to the tidal forcing,
we need to express $U(\br,t)$ in terms of
$Y_{lm}(\theta,\phi)$, the spherical harmonic function 
defined in the inertial frame centered on $M$ with the $z$-axis along 
${\hat{\bf S}}$.  This is achieved by the relation 
\be 
Y_{2m'}(\theta_L,\phi_L)=\sum_{m}\cD_\mmp (\Theta)
Y_{2m}(\theta,\phi), 
\ee 
where ${\cal D}_{mm'}(\Theta)$ is the Wigner ${\cal D}$-matrix of $l=2$
(e.g., Wybourne 1974), and we have chosen the $y$-axis along the direction 
${\hat{\bf S}}\times {\hat{\bf L}}$. The relevant $\cD_\mmp$'s are
\ba
&&\cD_{2\pm 2}={1\over 4}(1\pm\cosT)^2,\\
&&\cD_{2\pm 1}=-{1\over 2}\sinT (1\pm\cosT),\\
&&\cD_{20}={\sqrt{6}\over 4}\sin^2\!\Theta,\\
&&\cD_{10}=-{\sqrt{6}\over 2}\sinT \cosT,\\
&&\cD_{00}={1\over 2}(3\cos^2\!\Theta-1),
\ea
and
\be
\cD_{m'm}=(-1)^{m-m'}\cD_{mm'}=\cD_{-m,-m'}.
\ee
The tidal potential then becomes
\be
U(\br,t)=-\sum_\mmp U_\mmp\, r^2Y_{2m}(\theta,\phi)\,e^{-im'\Omega t},
\label{eq:poten}\ee
where 
\be
U_\mmp\equiv 
{GM'\over a^3}\cU_\mmp\equiv {GM'\over a^3}W_{2m'}\cD_\mmp(\Theta).
\ee
Note that when expressed in terms of $\phi_r$ (the azimuthal angle in 
the rotating frame of the star), each term in Eq.~(\ref{eq:poten}) 
has the dependence $e^{im\phi_r+im\Omega_s t-im'\Omega t}$.
Thus, the tidal potential is composed of various $(mm')$-components, 
each with forcing frequency $m'\Omega$ in the inertial frame.
In the frame corotating with the star, the forcing frequency of
the $(mm')$-component is $\tomega_\mmp=m'\Omega-m\Omega_s$ 
[Eq.~(\ref{eq:tomega})], with the pattern rotation frequency
\be
{\tomega_\mmp\over m}=\left({m'\over m}\right)\Omega-\Omega_s.
\ee
(Obviously, the pattern frequency has no meaning for the axisymmetric 
$m=0$ components.) Note that physically, there are 7 distinct components:
$(m,m')=(0,2),(\pm 1,2),(\pm 2,2),(1,0),(2,0)$ (see also Barker \& Ogilvie 2009).
The $(0,0)$ component 
is static and does not contribute to tidal dissipation. The
$(-m,-m')$-component is physically identical to the $(m,m')$-component.

\subsection{Ansatz for Tidal Response}

Each $(mm')$-component of the tidal potential drives fluid
perturbation inside the star, which can be specified by the
Lagrangian displacement $\bxi_\mmp(\br,t)$ and Eulerian density
perturbation $\delta\rho_\mmp(\br,t)$. In the absence of
dissipation, these perturbations are proportional to
$(U_\mmp/\omega_0^2) e^{im\phi-im'\Omega t}$ (where $\omega_0\equiv
\sqrt{GM/R^3}$ is the dynamical frequency of the star), exactly in
phase with the tidal potential. When fluid dissipation is present, 
there will be phase shift between the fluid perturbation and the 
tidal potential. This phase shift, in general, depends on $m$, the 
forcing frequency (in the rotating frame) $\tomega_\mmp$, as well 
as the intrinsic property (including the rotation rate) of the star. 
We write this phase shift as
\be
\Delta_\mmp=\tomega_\mmp t_\mmp,
\ee
where $t_\mmp$ is the ``lag time'' associated with the 
$(mm')$-perturbation. Thus, 
without loss of generality, we can write
$(mm')$-component of the fluid displacement as
\be
\bxi_\mmp(\br,t)={U_\mmp\over\omega_0^2}
\bar\bxi_\mmp(\br)\,
\exp(-im'\Omega t+i\Delta_\mmp).
\ee
The corresponding density perturbation is
\be
\delta\rho_\mmp(\br,t)={U_\mmp\over\omega_0^2}
\delta\bar\rho_\mmp(\br)\,
\exp(-im'\Omega t+i\Delta_\mmp),
\label{eq:deltarho}\ee
with
\be
\delta\bar\rho_\mmp=-\nabla\cdot (\rho\bar\bxi_\mmp).
\ee
Note that $\bar\bxi_\mmp(\br)$ and $\delta\bar\rho_\mmp(\br)$ 
are proportional to $e^{im\phi}$; except for this factor, they are 
real functions. Also note that 
\be
{U_{mm'}\over\omega_0^2}=
{M'\over M}\!\left({R\over a}\right)^3\cU_\mmp,
\ee
with $\cU_\mmp\equiv W_{2m'}\cD_\mmp(\Theta)$.
Thus, $\bar\bxi_\mmp(\br)$ and $\delta\bar\rho_\mmp(\br)$ 
specify the amplitudes (and shapes) the fluid perturbations after factoring 
out the ``equilibrium'' dimensionless tidal distortion, $(M'/M)(R/a)^3$.

Note that although our ansatz of the tidal responses is
motivated by the weak friction theory of
equilibrium tides, it is actually quite general, provided
that one keeps in mind that the lag time $t_\mmp$ depends
on $m$, $\tomega_\mmp$ and the intrinsic property of the star.

\subsection{Tidal Torque and Energy Transfer Rate}

The tidal torque on the star is
\be
{\bf T}=\int\!\!d^3\!x\,\delta\rho(\br,t)\,\br\times
\left[-\nabla U^\ast(\br,t)\right],
\ee
and the tidal energy transfer rate (from the orbit to the star) is
\be
\dot E=\int\!\!d^3\!x\,\rho(\br)\,{\partial\bxi(\br,t)\over \partial t}\cdot
\left[-\nabla U^\ast(\br,t)\right].
\ee
Using Eqs.~(\ref{eq:poten}) and (\ref{eq:deltarho}), the $z$-component 
(along the spin axis) of the tidal torque reduces to 
\be
T_z=T_0\sum_\mmp\cU_\mmp^2 m\,\kappa_\mmp\,\sin\!\Delta_\mmp,
\label{eq:Tz}\ee
where
\be
T_0\equiv G\left({M'\over a^3}\right)^2R^5,
\ee
and $\kappa_\mmp$ is the ``Love coefficient'':
\be
\kappa_\mmp={1\over MR^2}\int\!\!d^3\!x\,\delta\bar\rho_\mmp(\br)r^2
Y_{2m}^\ast(\theta,\phi).
\label{eq:love}\ee
Similaly, the energy transfer rate is given by
\be
\dot E=T_0\Omega\sum_\mmp\cU_\mmp^2 m'
\kappa_\mmp\,\sin\!\Delta_\mmp.
\label{eq:edot}\ee
Note that the $(mm')$-component of energy transfer and tidal torque
satisfies
\be
(\dot E)_\mmp={m'\Omega\over m}(T_z)_\mmp.
\ee
This is expected since the pattern rotation rate (in the inertial frame)
of the tidal force is $m'\Omega/m$.

Equations (\ref{eq:Tz}) and (\ref{eq:edot}) can be worked out explicitly.
Assuming $|\Delta_\mmp|\ll 1$, we have
\ba
&&T_z={(W_{22})^2\over 2}T_0 \Bigl[
(1+\ct)^4(\Omega-\Omega_s)\tau_{22}\nonumber \\
&&\qquad +\st^2(1+\ct)^2(2\Omega-\Omega_s)\tau_{12}\nonumber\\
&&\qquad -\st^2(1-\ct)^2(2\Omega+\Omega_s)\tau_{-12}\nonumber\\
&&\qquad -(1-\ct)^4(\Omega+\Omega_s)\tau_{-22}\Bigr]\nonumber\\
&&\qquad -{3(W_{20})^2}T_0\,\Omega_s\bigl(\st^4\,\tau_{20}
+\st^2\,\ct^2\,\tau_{10}\bigr).
\label{eq:Tz1}\ea
and
\ba
&&\dot E={(W_{22})^2\over 2}T_0\Omega \Bigl[
(1+\ct)^4(\Omega-\Omega_s)\tau_{22}\nonumber \\
&&\qquad +2\st^2(1+\ct)^2(2\Omega-\Omega_s)\tau_{12}
+6\,\st^4\,\Omega\,\tau_{02}\nonumber\\
&&\qquad +2\st^2(1-\ct)^2(2\Omega+\Omega_s)\tau_{-12}\nonumber\\
&&\qquad +(1-\ct)^4(\Omega+\Omega_s)\tau_{-22}\Bigr],
\label{eq:edot1}\ea
where we have defined
\be
\tau_\mmp=t_\mmp\kappa_\mmp,
\ee
and $\ct\equiv\cos\!\Theta$, $\st\equiv \sin\!\Theta$, and we have used
the identity $\tau_{-m,-m'}=\tau_{m,m'}$.

The perpendicular component of the tidal torque ${\bf T}$ 
is less straightforward to evaluate. The $y$-component
does not depend on tidal dissipation (for $|\Delta_\mmp|\ll 1$) and contributes 
to the spin precession. Since $T_y$ does change the spin-orbit elements, 
we will not consider it further.
The $x$-component of ${\bf T}$ is
\be
T_x=T_0\sum_{m\bm m'}\cU_{mm'}\cU_{\bm m'}\Delta_{mm'}
\kappa_{m\bm m'},
\label{eq:Tx}\ee
where
\ba
&&\kappa_{m\bm m'}\equiv {1\over i MR^2}\int\!\!d^3\!x\,\delta\bar\rho_\mmp(\br)\,r^2
\Bigl({\partial Y_{2\bm}^\ast\over\partial\theta}\sin\!\phi
\nonumber\\
&&\qquad\qquad\qquad -{i\bm\over\tan\theta}
Y_{2\bm}^\ast\cos\!\phi\Bigr).
\ea
Since $\delta\bar\rho_\mmp(\br)\propto e^{im\phi}$, $\kappa_{m\bm m'}$ is nonzero
only for $\bm=m\pm 1$. If we define
\be
\kappa_{mm'}^\pm\equiv\kappa_{m,\bm=m\pm 1,m'},
\ee
then Eq.~(\ref{eq:Tx}) can be written as
\be
T_x=T_0\sum_\mmp \cU_{mm'}\Delta_\mmp \left(\cU_{m+1,m'}\kappa_{mm'}^+
+\cU_{m-1,m'}\kappa_{mm'}^-\right).
\ee
A direct calculation shows that the coefficients $\kappa_{mm'}^\pm$ 
are related to the Love coefficients [Eq.~(\ref{eq:love})]
by
\ba
&&\kappa_{2m'}^-=\kappa_{2m'},\quad \kappa_{-2m'}^+=\kappa_{-2m'},\nonumber\\
&&\kappa_{1m'}^+=\kappa_{1m'},\quad \kappa_{1m'}^-
=\sqrt{3/2}\,\kappa_{1m'},\nonumber\\
&&\kappa_{0m'}^\pm=\sqrt{3/2}\,\kappa_{0m'},\nonumber\\
&&\kappa_{-1m'}^+=\sqrt{3/2}\,\kappa_{-1m'},\quad \kappa_{-1m'}^-=\kappa_{-1m'}
\ea
Thus $T_x$ can be worked out explicitly:
\ba
&&T_x={(W_{22})^2\over 2}T_0 \Bigl[
\st(1+\ct)^3(\Omega-\Omega_s)\tau_{22}\nonumber \\
&&\qquad +\st(1+\ct)^2(2-\ct)(2\Omega-\Omega_s)\tau_{12}\nonumber\\
&&\qquad +6\,\st^3\,\Omega\, \tau_{02}\nonumber\\
&&\qquad +\st(1-\ct)^2(2+\ct)(2\Omega+\Omega_s)\tau_{-12}\nonumber\\
&&\qquad +\st(1-\ct)^3(\Omega+\Omega_s)\tau_{-22}\Bigr]\nonumber\\
&&\qquad +{3(W_{20})^2}\,T_0 \,\Omega_s\,
\bigl(\st^3\,\ct\,\tau_{20}+\st\,\ct^3\,\tau_{10}\bigr).
\label{eq:Tx1}\ea

\subsection{Tidal Evolution Equations}

Given the tidal torque $T_z$ and energy transfer rate $\dot E$, 
the tidal evolution equations for the stellar spin $\Omega_s$ and
the orbital semi-major axis $a$ are
\be
\dot\Omega_s={T_z\over I},\qquad
{\dot a\over a}=-{2a{\dot E}\over GMM'},
\label{eq:dotOmegas}\ee
where $I$ is the moment of the inertia of the star. 
The spin-orbit misalignment angle
$\Theta$ is given by $\cos\Theta={\bf S}\cdot {\bf L}/(SL)$, where
$S=I\Omega_s$ and $L=\mu a^2\Omega$. Using 
${\dot{\bf S}}=-{\dot{\bf L}}={\bf T}$, we find
\be
\dot\Theta=-{N_x\over S}-{N_x\over L} \cosT+{N_z\over L} \sinT.
\label{eq:dotTheta}\ee
Note that the rate of change for the magnitude of 
the orbital angular momentum satisfies
\be
\dot L=-T_z\cosT-T_x\sinT=-{\dot E\over\Omega}={\dot E_{\rm orb}\over
\Omega}.
\ee
This can be checked directly using Eqs.~(\ref{eq:Tz1}), (\ref{eq:edot1})
and (\ref{eq:Tx1}). Thus a circular orbit will remain circular, as it should be.

These evolution equations for $\dot\Omega_s$, $\dot a$ and $\dot\Theta$
[with $T_z,~{\dot E},~T_x$ given by Eqs.~(\ref{eq:Tz1}),~(\ref{eq:edot1}),~
(\ref{eq:Tx1}), respectively] are the most general tidal equations for
circular binaries. They depend on 7 independent ``reduced'' tidal lag 
times $\tau_{mm'}=t_{mm'}\kappa_{mm'}$, corresponding to the 7 
independent tidal forcing components. In general, each $\tau_{mm'}$
depends on $m$, $\tomega_{mm'}$ and the intrinsic property (including
$\Omega_s$) of the star.

\subsection{Special Case: Weak Friction Theory of Equilibrium Tide}

When $\tau_{mm'}=\tau$ are the same for all seven tidal forcing
components, we find
\ba
&&\dot E={12\pi\over 5}T_0\Omega \Bigl(\Omega-\Omega_s\cosT\Bigr)\tau,\\
&&T_z={6\pi\over 5}T_0\left[2\Omega\cosT-\left(1+\cos^2\!\Theta\right)
\Omega_s\right]\tau,\\
&&T_x={6\pi\over 5}T_0\sinT \Bigl(2\Omega-\Omega_s\cosT\Bigr)\tau.
\ea
These give
\ba
&&{\dot a\over a}=-{1\over t_a}\left(1-{\Omega_s\over\Omega}\cosT
\right),\label{eq:adoteq}\\
&&{\dot\Omega_s\over\Omega_s}={1\over t_a}\left({L\over 2S}\right)
\left[\cos\Theta-\left({\Omega_s\over 2\Omega}\right)(1+\cos^2\Theta)
\right],\\
&&\dot\Theta=-{1\over t_a}\left({L\over 2S}\right)\sin\Theta
\left[1-\left({\Omega_s\over 2\Omega}\right)\left(\cosT-{S\over L}\right)\right],
\label{eq:thetadoteq}\ea
where
\be
{1\over t_a}\equiv {3k_2\over Q}\left({M'\over M}\right)
\left({R\over a}\right)^5\Omega,
\ee
and we have defined 
\be
k_2 \Delta t_L\equiv {4\pi\over 5}\tau,\quad 
Q\equiv \bigl(2\Omega\Delta t_L\bigr)^{-1}.
\label{eq:defk2}\ee
Here $k_2$ and $\Delta t_L$ have the usual meanings as in the
equilibirum tide theory: $k_2$ is the Love number and $\Delta t_L$ is
the tidal lag time [which is related to the viscous time $t_{\rm vis}$
by $\Delta t_L=1/(\omega_0^2 t_{\rm vis})$.]
These equations agree with those given in Alexander (1973) and others
(Hut 1980; Eggleton et al.~1998).


\section{Tidal Torque due to Inertial Wave Dissipation}

The convective envelope of a solar-type star supports inertial waves,
which are driven entirely by Coriolis force.  The frequency $\tomega$
(in the rotating frame) of an inertial wave is related to its (local)
wavenumber vector $\bk$ by the dispersion relation
(e.g., Greenspan 1968)
\be
\tomega^2=\left(2\bOmega_s \cdot \bk/|\bk|\right)^2.
\label{eq:dispersion}\ee
Thus inertial waves exist only when $|\tomega|<2\Omega_s$.

In hot Jupiter systems with $\Omega\gg \Omega_s$, the only tidal
forcing component that is capable of exciting inertial waves in the
star is $(m,m')=(1,0)$, with the forcing frequency 
(in the rotating frame)
$\tomega=-\Omega_s$. [The $(-1,0)$-component has $\tomega=\Omega_s$
and is physically identical.]
The tidal torque associated with this component can be read off
directly from Eqs.~(\ref{eq:Tz1}) and (\ref{eq:Tx1}):
\ba
&&\left(T_z\right)_{10}=-{3\pi\over 5}T_0\Omega_s \tau_{10}\,\left(\sin\!\Theta
\cos\!\Theta\right)^2,\label{eq:Tz3}\\
&&\left(T_x\right)_{10}={3\pi\over 5}T_0\Omega_s \tau_{10}\,\sin\!\Theta\cos^3\!\Theta,
\ea
with $T_0=G(M'/a^3)^2R^5$. Similar to Eq.~(\ref{eq:defk2}), we define
the relevant tidal Love number $k_{10}$, lag time $\Delta t_{10}$ and
quality factor $Q_{10}$:
\be
k_{10}\Delta t_{10}\equiv {4\pi\over 5}\tau_{10},\quad
Q_{10}\equiv \left(\Omega_s\Delta t_{10}\right)^{-1}.
\label{eq:Q10}\ee
Using Eqs.~(\ref{eq:dotOmegas}) and (\ref{eq:dotTheta}), we find
\ba
&&\left({\dot\Omega_s\over\Omega_s}\right)_{10}
=-{1\over t_{s10}}\left(\sinT\cosT\right)^2,\label{eq:omegdot}\\
&&\left(\dot\Theta\right)_{10}=-{1\over t_{s10}}\,\sinT
\cos^2\!\Theta\left(\cosT+{S\over L}\right),
\label{eq:dTheta10}\ea
where
\ba
&&{1\over t_{s10}}={3\pi\tau_{10} T_0\over 5I}\nonumber\\
&&\qquad ={3k_{10}\over 4Q_{10}}\left({M'\over M}\right)
\left({R\over a}\right)^5\left({L\over S}\right)\Omega.
\ea
For $M=M_\star$ and $M'=M_p$, the corresponding timescale is 
\ba
&& t_{s10}=4.3\,\left({\kappa\over 0.1}\right)\left({Q_{10}/k_{10}\over 10^7}\right)
\left({M_\star\over 10^3M_p}\right)^2\left({\bar\rho_\star\over\bar\rho_\odot}\right)
\nonumber\\
&&\qquad ~~~ \times \left({10\,{\rm d}\over P_s}\right)
\left({P\over 1\,{\rm d}}\right)^4\,{\rm Gyr}
\ea
[cf.~Eq.~(\ref{eq:tTheta})].

Since the $(1,0)$-component of the tidal potential is static
in the inertial frame, the associated energy transfer rate $\dot E$ 
associated with this component is zero, giving 
\be 
\left(\dot a\right)_{10}=0.
\ee
This does not imply zero tidal 
dissipation. In fact, the tidal energy dissipation rate 
equals the energy transfer rate in the rotating frame,
\be
\dot E_r=-\Omega_s T_z>0.
\ee
This is exactly balanced by $(d/dt)(S^2/2I)=\Omega_s \dot S=\Omega_s T_z<0$,
so that $\dot E=\dot E_r+\Omega_s T_z=0$. Also note that 
the $(1,0)$-tidal force does not change the magnitude of the orbital 
angular momentum, $\dot L=-T_z\cosT-T_x\sinT=0$. So a circular orbit will 
remain circular.

Two interesting features of Eq.~(\ref{eq:dTheta10})
are worth noting: (i) $\dot\Theta=0$ when
$\Theta=90^\circ$. Thus, if the $(1,0)$-tidal component is the only
tidal force operating in the system, there could be many systems with
the spin-orbit misalignment angle stalled around $90^\circ$.  (ii)
$\dot\Theta>0$ when $\cos\!\Theta < -S/L$. This implies that a
retrograde system ($\Theta>\pi/2$) may evolve toward anti-alignment.
Of course, other tidal components can also contribute to $\dot\Theta$
(Sect~.2) and will weaken these features. But with enough statistics
of misaligned systems, it may be possible to test these or constrain
$k_{10}/Q_{10}$.

A detailed calculation of $Q_{10}/k_{10}$ is beyond the scope of this
paper. Previous works of tidal dissipation in rotating solar-type
stars have focused on the $m=2$ tide for aligned binaries (Savonije \&
Papaloizou 1997; Papaloizou \& Savoniji 1997; Savonije \& Witte 2002;
Ogilvie \& Lin 2007) and demonstrated the importance of 
inertial waves. In the numerical study by Ogilvie \& Lin (2007),
inertial waves affect tidal dissipation in two ways: 
(i) short-wavelength inertial waves in the
convection zone are damped by turbulent viscosity; (ii) Inertial waves
influences the excitation of gravity (Hough) waves in the radiative zone.
Regarding (ii), they assumed that the inward propagating Hough waves are 
damped near the stellar center; this is appropriate for binary stars,
but may lead to an over-estimate of stellar 
tidal dissipation in hot Jupiter systems (Barker \& Ogilvie 2010,2011;
see Sect.~1.2). Ogilvie \& Lin (2007) found that when inertial
waves are excited ($|\tomega|<2\Omega_s$), the tidal dissipation 
rate depends on $\tomega$ in an erratic manner. On average, the energy 
dissipation rate in the convection zone is significantly increased 
(by 1-3 orders of magntitude, depending on the spin period) 
compared to equilibrium tides. 
For example, Figure 3-6 of Ogilvie \& Lin (2007) show that for a solar-type star
at $\tomega=-\Omega_s$, the tidal quality factor associated with 
inertial wave dissipation is $Q_\star'\simeq 6\times 10^7$ for $P_s=10$~days
and $Q_\star'\simeq 5\times 10^6$ for $P_s=3$~days.

Barker \& Ogilvie (2009) reported the result of a calculation of the
$l=2,\,m=1$ tidal dissipation for a specific F-type stellar model
(appropriate for the hot Jupiter system XO-3) using the numerical
method of Ogilvie \& Lin (2007). Intriguingly, they found that at
$\tomega/\Omega_s=-1$, tidal dissipation is significantly enhanced,
with the effective tidal $Q\sim 10^6$ (see their Fig.~7). They suggested
that this prominent feature arises from 
resonant excitation of the $l=m=1$ Rossby mode. More systematic studies
on the $(m,m')=(1,0)$ tide for a range of stellar models
(with different sizes of the convective envelope) would be useful.


\section{Conclusion}

The main point of this paper is that in close-in exoplanetary
systems, when the stellar spin axis is misaligned with the orbital
angular momentum axis, a new tidal dissipation channel opens up. This
channel involves the excitation of inertial waves in the stellar
convection zone, and is forbidden for aligned systems.  Thus, tidal
damping of spin-orbit misalignment can be more efficient than orbital
decay. This may explain the stellar obliquity -- effective temperature
correlation observed by Winn et al.~(2010) and the obliquity -- age
correlation noted by Triaud (2011), while still
being consistent with the survival of hot Jupiters with very short
orbital period (see Sect.~1.1).

On a more general level, this paper highlights the importance
of treating tidal dissipation as being dependent on the tidal 
forcing frequency, the strength of tidal potential, and the
tidal processes involved (e.g., orbital decay vs. spin-orbit alignment),
in contrast to the equilibrium tide equations widely used
in many applications and empirical works (see references in 
Sect.~1.2). Indeed, various studies of the physical
mechanisms of tidal dissipation (e.g., Ogilvie \& Lin 2004,2007; 
Goodman \& Lackner 2009; Barker \& Ogilvie 2010,2011; 
see also Zahn 2008 for a review of earlier works) have already made this 
point clear. While it is recognized that the weak friction theory of 
equilibrium tides (Darwin 1880; Goldreich \& Soter 1966; Alexander 1973;
Hut 1981; Eggleton et al.~1998) is a parameterized theory (with the tidal $Q$ or
lag time being the single parameter), we have shown in this paper that 
even at the parameterized level, its equations are sometimes inadequate
or misleading, since different tidal processes (e.g., orbital
decay vs. spin-orbit alignment) may involve very different tidal dissipation
mechanisms.

Given the complicated nature of tidal dissipation, a parameterized
``effective'' theory of tidal evolution remains useful. We have
derived such an effective theory in this paper (see Sect.~2). For
misaligned circular binaries, there are seven independent tidal
quality factors $Q_{mm'}$ or lag times $\tau_{mm'}$ in the theory,
related to the tidal responses for different Fourier components of the
tidal potential. Obviously, such a system of equations with 7 parameters 
(Sect.~2.4) is not convenient to use in real applications. For hot 
Jupiter systems, we suggest that a reduced system of equations involving two
tidal $Q$ parameters may be adopted. They are
\ba
&&{\dot a\over a}=\left({\dot a\over a}\right)_{\rm eq},\\
&&{\dot\Omega_s\over\Omega_s}=\left({\dot\Omega_s\over\Omega_s}\right)_{\rm eq}+
\left({\dot\Omega_s\over\Omega_s}\right)_{10}
-\left({\dot\Omega_s\over\Omega_s}\right)_{10,{\rm eq}},\\
&&\dot\Theta=(\dot\Theta)_{\rm eq}+(\dot\Theta)_{10}
-(\dot\Theta)_{10,{\rm eq}},
\ea
where $(\dot a/a)_{\rm eq}$, $(\dot\Omega_s/\Omega_s)_{\rm eq}$
and $(\dot\Theta)_{\rm eq}$ are given by 
Eqs.~(\ref{eq:adoteq})-(\ref{eq:thetadoteq}) and characterized by
the parameter $k_2/Q$, while $(\dot\Omega_s/\Omega_s)_{10}$
and $(\dot\Theta)_{10}$ are given by Eqs.~(\ref{eq:omegdot})-(\ref{eq:dTheta10})
and characterized by the parameter $k_{10}/Q_{10}$.
The expressions for $(\dot\Omega_s/\Omega_s)_{10,{\rm eq}}$ and
$(\dot\Theta)_{10,{\rm eq}}$ are the same as 
Eqs.~(\ref{eq:omegdot})-(\ref{eq:dTheta10}), except that the 
the parameter $k_{10}/Q_{10}$ should be replaced by $k_2/Q$.
These equations (extended to eccentric orbits), combined with similar equations
for planetary tides, can be used to assess and constrain the effects
of tidal evolution in close-in exoplanetary systems.

\section*{Acknowledgments}
I thank Josh Winn for update on current observations and useful
discussion, and Gordon Ogilvie for helpful communication and comment.
This work has been supported in part by AST-1008245.


\end{document}